\documentclass{appolbhep}
\usepackage{epsfig}

\usepackage{amssymb}

\def\be{\begin{equation}}
\def\ee{\end{equation}}

\begin{document}
\title{Diffraction and
\boldmath{$\sigma_{\gamma^* p}$}
\thanks{Presented by D.~Schildknecht at DIS2002, Cracow, Poland, 30 April to 4 May 2002 \hfill \\
Supported by the BMBF, Contract 05 HT9BBA2  }%
}
\author{D.~Schildknecht and M.~Tentyukov
\address{Fakult{\"a}t f{\"u}r Physik, Universit\"at Bielefeld,\\
Universit{\"a}tsstra{\ss}e 25, D-33615 Bielefeld, Germany   }
\and
M.~Kuroda
\address{Department  of
Physics, Meiji-Gakuin
University\\ Yokohama 244, Japan }
\and
B.~Surrow
\address{
Brookhaven National Laboratory\\
Upton, NY 11973-5000 USA }
}
\maketitle
\begin{abstract}
The empirical scaling law, wherein the total photoabsorption cross section
depends on the single variable $\eta=(Q^2 + m^2_0)/\Lambda^2 (W^2)$, provides
empirical evidence for saturation in the sense of $\sigma_{\gamma^* p}
(W^2 , Q^2) /
\sigma_{\gamma p} (W^2) \rightarrow 1$ for $W^2 \rightarrow \infty$ at fixed
$Q^2$. The total photoabsorption
cross section is related to elastic diffraction in terms of a sum rule. The
excess of diffractive production over the elastic component is due to
inelastic diffraction that contains the production of hadronic states of higher
spins. Motivated by the diffractive mass spectrum, the generalized vector
dominance/color dipole picture (GVD/CDP)
is extended to successfully describe the DIS data in the
full region of $x \le 0.1$, all $Q^2 \ge 0$, where the diffractive
two-gluon-exchange mechanism dominates.
\end{abstract}

In the present talk, I wish to concentrate on the relation between the
total photoabsorption cross section, $\sigma_{\gamma^* p} (W^2 , Q^2)$,
at low $x \cong Q^2 / W^2 \le 0.1$ and diffractive production, $\gamma^* p
\rightarrow Xp$ \cite{1}.

The experimental data \cite{2,3,4,5,6}
on $\sigma_{\gamma^* p} (W^2 , Q^2)$ at $x \le 0.1$ and all $Q^2 \ge 0$,
including photoproduction $(Q^2 = 0)$, lie on a single curve \cite{7},
\be
\sigma_{\gamma^* p} (W^2, Q^2) = \sigma_{\gamma^*p} (\eta (W^2, Q^2)),
\label{(1)}
\ee
if plotted against the low-x scaling variable

\vspace{-0.4cm}
\be
\eta(W^2, Q^2) = \frac{Q^2 + m^2_0}{\Lambda^2 (W^2)},
\label{(2)}
\ee

\vspace{-0.3cm}\noindent
where $\Lambda^2(W^2)$ is a slowly increasing function of $W^2$ and $m^2_0 \cong
0.16\,{\rm GeV}^2$. Compare fig.~1
for a plot of $\sigma_{\gamma^* p} (W^2 , Q^2)$ against $\eta$. The
function $\Lambda^2 (W^2)$ may be represented, alternatively, by a power law
or by a logarithm,

\vspace{-0.3cm}
\be
\Lambda^2 (W^2) = \left\{ \matrix{ & C_1 (W^2 + W^2_0)^{C_2} , \cr
  & C^\prime_1 \ln \left( \frac{W^2}{W^2_0} + C^\prime_2 \right).\cr}
\right.
\label{(3)}
\ee

\begin{figure}[h]
\setlength{\unitlength}{1cm}
\begin{minipage}[t]{6.5cm}
\vspace*{1.5cm}
\begin{picture}(3.5,3.5)\psfig{file=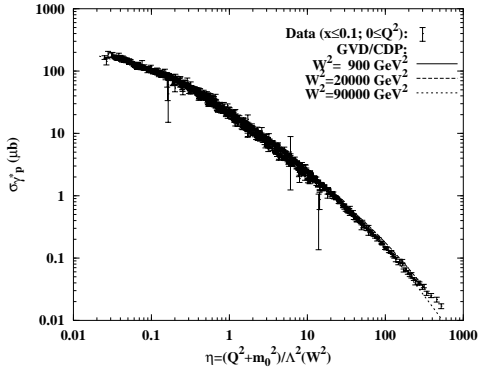,width=6.5cm}
\end{picture}\par
\end{minipage}\hfill
\begin{minipage}[t]{5.5cm}
\caption{
The experimental data \cite{2,3,4,5,6} for the total photoabsorption cross section,
$\sigma_{\gamma^* p} (W^2 , Q^2)$ as a function of $\eta (W^2 , Q^2)$
compared with the predictions from the GVD/CDP.}
\end{minipage}
\end{figure}

\vspace{-0.2cm}\noindent
We refer to refs.\cite{7,8} for the numerical values of the fit parameters in
$\Lambda^2 (W^2)$.

The empirical model-independent finding (\ref{(1)}) is interpreted in the
generalized vector
dominance/color dipole picture (GVD/CDP) \cite{7,8}
that rests on the generic structure of the two-gluon-exchange
virtual-photon-forward-Compton-scattering amplitude.
Evaluation of this amplitude in the $x\rightarrow 0$ limit and transition to
transverse position space implies \cite{9}

\vspace{-0.5cm}
\be
\sigma_{\gamma^*_{T,L} p} (W^2 , Q^2) =
\!\!\!\int\!\!\! dz \!\!\!\int\!\!\! d^2 r_\bot \!\!\!\sum_{\lambda,
\lambda^\prime = \pm 1}
| \psi_{T,L}^{(\lambda , \lambda^\prime)} ( \vec r_
\bot , z, Q^2) |^2
\sigma_{(q \bar q)p} (\vec r_\bot^{~2} , z, W^2),
\label{(4)}
\ee

\vspace{-0.3cm}\noindent
where the Fourier representation of the color-dipole cross section,

\vspace{-0.4cm}
\begin{eqnarray}
\sigma_{(q \bar q)p}  (\vec r_\bot^{~2} , z, W^2) &=& \int d^2 l_\bot
\tilde\sigma_{
(q \bar q)p} (\vec l^{~2}_\bot , z, W^2)(1 - e^{-i\vec l_\bot \vec r_\bot})
\label{(5)} \\
& = & \sigma^{(\infty)} \left\{ \matrix{ \frac{1}{4} \vec r^{~2}_\bot \langle
\vec l^{~2}_\bot \rangle_{W^2 ,z} , & {\rm for}~ \vec r^{~2}_\bot
\langle \vec l^{~2}_\bot \rangle_{W^2 , z} \rightarrow 0 , \cr
1, ~~~~~~
&\,\, {\rm for}~ \vec r^{~2}_\bot \langle \vec l^{~2}_\bot \rangle_{W^2 , z}
\rightarrow \infty , \cr } \right. \nonumber
\end{eqnarray}

\vspace{-0.2cm}\noindent
contains ``color transparency'' in the limit of $\vec r^{~2}_\bot
\langle \vec l^{~2}_\bot \rangle_{W^2 ,z} \rightarrow 0$, as well as hadronic
unitarity, provided

\vspace{-0.4cm}
\be
\sigma^{(\infty)} \equiv \pi \int \, d \vec l^{~2}_\bot \tilde\sigma (\vec
l^{~2}_\bot , z , W^2)
\label{(6)}
\ee

\vspace{-0.3cm}\noindent
has decent high-energy behavior. The average or effective gluon transverse
momentum, $\langle \vec l^{~2}_\bot \rangle_{W^2 ,z}$, in (\ref{(5)}) is given
by

\vspace{-0.2cm}
\be
\langle \vec l^{~2}_\bot \rangle_{W^2 ,z} \equiv \frac{\int d \vec l^{~2}_\bot
\vec l^{~2}_\bot \tilde\sigma_{(q \bar q)p} ( \vec l^{~2}_\bot, z, W^2)}
{\int d l^2_\bot \tilde\sigma_{(q \bar q)p} (\vec l^{~2}_\bot , z, W^2)}.
\label{(7)}
\ee

\vspace{-0.1cm}\noindent
It is a characteristic feature of the $x \rightarrow 0$ limit of the
two-gluon-exchange amplitude that the representation (\ref{(4)}) factorizes into
the product of the photon wave function, $|\psi |^2$, that describes the photon
coupling to the $q \bar q$ state and its propagation, and the
color-dipole cross section, $\sigma_{(q \bar q)p}$,
that describes the forward scattering of the color dipole from the proton. The
scattering is ``diagonal'' in the variables $\vec r , z$, since these variables
remain fixed during the scattering process.

The empirical scaling law (\ref{(1)}) is embodied in the representation
(\ref{(4)}) by requiring the dipole cross section (\ref{(5)}) to depend on
the product $\vec r^{~2} \cdot \Lambda^2 (W^2)$.
This implies that $\langle \vec l^{~2}_\bot \rangle_
{W^2 ,z}$ be proportional to $\Lambda^2 (W^2)$.
In the GVD/CDP,
we approximate the distribution in the gluon transverse momentum,
$\vec l^{~2}$, in (\ref{(5)}) by a $\delta$-function situated at the
effective gluon transverse momentum, $\langle \vec l^{~2}_\bot \rangle_{W^2 ,z}$,

\vspace{-0.5cm}
\be
\tilde\sigma_{(q \bar q) p} (\vec l^{~2}_\bot , z (1-z), W^2) = \sigma^{(\infty)}
\frac{1}{\pi} \delta (\vec l^{~2}_\bot - \Lambda^2 (W^2) z (1 - z)).
\label{(8)}
\ee
The proportionality factor $z (1-z)$ in (\ref{(8)})
is a model assumption that improves the high-$Q^2$ behavior. With
(\ref{(5)}) and (\ref{(8)}), and the Fourier representation of the wave function
inserted, the expression for the cross section (\ref{(4)}) may be evaluated
analytically in
momentum space \cite{7}. We only note the approximate final expression

\vspace{-0.5cm}
\be
\sigma_{\gamma^* p} (W^2 , Q^2) = \frac{\alpha R_{e^+ e^-}}{3\pi}
\sigma^{(\infty)}
\left\{ \matrix{ \ln (1/\eta) , & {\rm for}\, \eta\rightarrow
m^2_0/\Lambda^2 (W^2) , \cr
1 / 2\eta, & {\rm for} \, \eta \gg 1 . \cr } \right.
\label{(9)}
\ee

\vspace{-0.3cm}\noindent
and refer to ref.\cite{7} for details.

According to (\ref{(9)}), at any fixed value of $Q^2$, for sufficiently large
$W$, a soft, logarithmic energy dependence is reached for $\sigma_{\gamma^* p}$.
The GVD/CDP that rests on the generic structure of the two-gluon exchange from
QCD, and contains hadronic unitarity and scaling in $\eta$, leads to the
important conclusion that

\vspace{-0.5cm}
\be
\lim_{{W^2 \rightarrow \infty}\atop{Q^2 = {\rm const}}} \, \frac{\sigma_{\gamma^*
p} (W^2 , Q^2)}{\sigma_{\gamma p} (W^2)} = 1.
\label{(10)}
\ee

\vspace{-0.1cm}\noindent
The behavior (\ref{(9)}) may be called ``saturation''. Since the low x (HERA)
data, according to fig.~1, show evidence for the behavior (\ref{(9)})
that implies (\ref{(10)}), we may
indeed conclude that HERA yields evidence for ``saturation''. Needless to stress,
future tests of scaling in $\eta$, by increasing $W$ as much as possible, are
clearly desirable to provide further evidence for the validity of the
remarkable conclusion (\ref{(10)}) that puts virtual and real
photoproduction on equal footing at any fixed $Q^2$ in the limit of infinite
energy.

We turn to diffractive production. The diagonal form (\ref{(4)}) of
$\sigma_{\gamma^*_{T,L}p}$, or rather of the virtual forward-Compton-scattering
amplitude, develops its full power when considering diffractive production,
$\gamma^* p \rightarrow Xp$.
The two-gluon-exchange generic structure for $x \rightarrow 0$ implies \cite{10}

\vspace{-0.5cm}
\begin{eqnarray}
& & \frac{d\sigma_{\gamma^*_{T,L}p\rightarrow X p}(W^2, Q^2, t)}{dt}
\Bigg|_{t=0} \label{(11)} \\
& & = \frac{1}{16\pi} \int^1_0 dz
\int d^2 r_\bot \sum_{\lambda, \lambda^\prime = \pm 1} |
\psi_{T,L}^{(\lambda, \lambda^\prime)}(r_\bot , z, Q^2) |^2
\sigma_{(q \bar q)p}^2 (\vec r^{~2}_\bot , z, W^2) .
\nonumber
\end{eqnarray}

\vspace{-0.3cm}\noindent
Note the close analogy of (\ref{(11)}) to the simple $\rho^0$ dominance formula
for photoproduction \cite{11}

\vspace{-0.5cm}
\be
\frac{d\sigma}{dt} \Bigg|_{t=0} (\gamma p \rightarrow \rho^0 p) = \frac{1}
{16\pi} \frac{\alpha \pi}{\gamma^2_\rho} \sigma^2_{\rho^0 p} .
\label{(12)}
\ee

\vspace{-0.2cm}\noindent
Upon transition to the momentum-space representation in (\ref{(11)}) and after
integration over all variables with the exception of the mass $M$ of the
outgoing state $X$, one obtains the mass spectrum,
$d\sigma_{\gamma^*_{T,L}p \rightarrow Xp} / dtdM^2$ for forward
production that depends on $W^2 , Q^2$ and $M^2$.
A comparison of this mass spectrum with the integrand of the total cross
section in (\ref{(4)}) (obtained upon transition to momentum space and
appropriate integration with the exception of one final integration over
$M^2$), allows one to rewrite (\ref{(4)}) as a sum rule that reads \cite{1}

\vspace{-0.6cm}
\begin{eqnarray}
& & \sigma_{\gamma^* p} (W^2, Q^2) = \sqrt{16\pi}
\sqrt{\frac{\alpha R_{e^+e^-}}{3 \pi}} \label{(13)} \\
& & \cdot \int_{m^2_0} dM^2
\frac{M}{Q^2 + M^2} \left[\sqrt{\frac{d\sigma_{\gamma^*_T}}{dt dM^2} \Bigg|_{t=0}}
+ \sqrt{\frac{Q^2}{M^2}} \sqrt{\frac{d\sigma_{\gamma^*_L}}{dt dM^2} \Bigg|_{t=0}}
\,\,\right] .\nonumber
\end{eqnarray}

\vspace{-0.3cm}\noindent
The sum rule represents the total photoabsorption cross section in terms of
diffractive forward production. It is amusing to note that (\ref{(13)}) is
the virtual-photon analogue of the photoproduction sum rule \cite{11}

\vspace{-0.4cm}
\be
\sigma_{\gamma p}(W^2) = \sum_{V= \rho^0 , \omega , \phi, ...}
\sqrt{16\pi} \sqrt{\frac{\alpha\pi}{\gamma^2_V}}
\sqrt{\frac{d\sigma_{\gamma p\rightarrow V_0}}{dt} \Bigg|_{t=0}}.
\label{(14)}
\ee

\vspace{-0.2cm}\noindent
based on $\rho^0, \omega , \phi$ dominance. Note, however, that (\ref{(13)}) is
a strict consequence of the generic two-gluon exchange structure evaluated in the
$x \rightarrow 0$ limit that forms the basis of the GVD/CDP\footnote{The sum
rule (\ref{(13)}) is also obtained from GVD arguments by themselves \cite{Spie}}.

It is evident, even though apparently always ignored, that the diffractive
production cross section (\ref{(11)}) describes elastic and only elastic
diffraction, where ``elastic'' is meant to denote diffractive production of
hadronic states $X$ that carry photon quantum numbers. Otherwise, the color
dipole cross section under the integral in (\ref{(11)}) could never be
identical to the one in
(\ref{(4)}), and (\ref{(13)}) could never follow from (\ref{(4)}) and
(\ref{(11)}).

``Inelastic'' diffraction, namely diffractive production of states with spins
different from the projectile spin, subject to the restriction of natural
parity exchange, is a well-known phenomenon in hadron physics \cite{Stod}.
Evidence for
inelastic diffraction in DIS is provided by the decrease \cite{12}
of the average thrust angle (``alignment'') with increasing mass of the
produced state $X$. This observation implies production of hadronic states $X$
that do not exclusively carry photon quantum numbers.

It is, accordingly, not surprising that the elastic diffraction obtained
from (\ref{(11)}) with the parameters employed for $\sigma_{\gamma^* p}$
underestimates the measured cross section considerably, in particular for high
values of the mass $M$ of the state $X$. Compare fig.~2 taken from ref.\cite{1}.

\begin{figure}[h]
\setlength{\unitlength}{1cm}
\begin{minipage}[t]{6.5cm}
\vspace*{1.7cm}
\begin{picture}(3.5,3.5)\epsfig{file=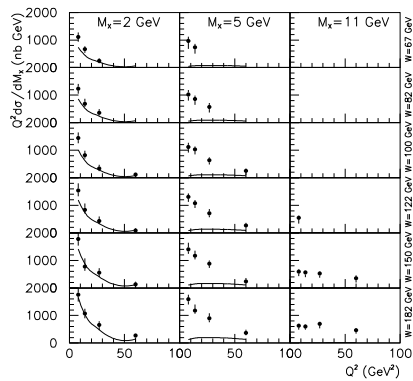,width=6.0cm}
\end{picture}\par
\end{minipage}\hfill
\begin{minipage}[t]{5.5cm}
\caption{
The cross section for elastic diffractive production (GVD/CDP) as a function of $Q^2$
compared with ZEUS data from ref. \cite{ZEUS}.
}
\end{minipage}
\vspace*{-0.5cm}
\end{figure}

Theoretical approaches \cite{13,14,15,16} to the description of high-mass
diffractive production frequently introduce a quark-antiquark-gluon
$(q \bar q g)$ component in the incoming photon. As this component is usually
ignored \cite{13,14,15} in the treatment of the total cross section,
I am afraid, there
is the danger of an inconsistency, due to a violation of the optical theorem.
A consistent inclusion of the
$q \bar q g$ component in elastic diffraction is contained in ref.\cite{16},
while an attempt for a
consistent and unified treatment of inelastic and elastic diffraction and
the total cross section, is provided in ref.\cite{17}.

I return to the analysis of the total cross section. The above discussion of
diffraction, in particular the sum rule (\ref{(13)}), suggests to introduce an
upper limit \cite{1} in the integration over $dM^2$ in $\sigma_{\gamma^* p}
(W^2 , Q^2)$. At finite energy $W$, the diffractively produced mass spectrum is
undoubtedly bounded by an upper limit that increases with energy. In our previous
analysis \cite{7,8}, we ignored such an upper limit, since the contribution
of high
masses seemed to be suppressed anyway. We have examined the effect of a
cut-off, $m^2_1$, in the momentum space version of (\ref{(4)}) or, equivalently,
in (\ref{(13)}). Putting

\vspace{-0.3cm}
\be
m^2_1 = (22 \, {\rm GeV})^2 = 484 \, {\rm GeV}^2 ,
\label{(15)}
\ee

\vspace{-0.1cm}\noindent
that is the mass of the largest bin
in the ZEUS data \cite{12}, we obtain an excellent description of all data with
$x \le 0.1$, all $Q^2 \ge 0$, as shown in fig.~1. Putting $m^2_1 = \infty$
overestimates the cross section $\sigma_{\gamma^* p}$ significantly for
$\eta \ge 10$, while values of $m^2_1$ smaller than the upper bound (\ref{(15)})
yield results below the experimental
ones at large $\eta$.
It is gratifying that the simple procedure of introducing a
cut-off\footnote{The simple cut-off procedure leads to a small violation of
scaling in $\eta$ for $\eta \geq 50$ (compare fig.~1) that may presumably be
avoided by a refined treatment.}
 that
(aproximately) coincides with the upper limit for diffractive production extends
the GVD/CDP to the full region of $x \le 0.1$, all $Q^2 \ge 0$, where
diffraction dominates the virtual Compton-forward-scattering amplitude.

\begin{figure}[h]
\setlength{\unitlength}{1cm}
\begin{minipage}[t]{6.5cm}
\vspace*{1.0cm}
\begin{picture}(3.5,3.5)\psfig{file=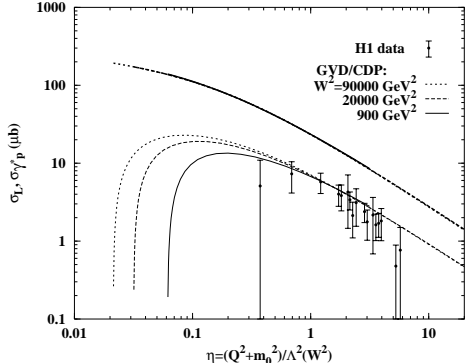,width=6.5cm}
\end{picture}\par
\end{minipage}\hfill
\begin{minipage}[t]{5.5cm}
\caption{
The longitudinal part of the photoabsorption cross section as a function
of $\eta$ compared with H1 data.
}
\end{minipage}
\vspace*{-0.5cm}
\end{figure}

In fig.~3 we show the prediction \cite{18} of the GVD/CDP for the longitudinal
cross section in comparison with data from an H1 analysis \cite{19}.

In conclusion:\\
i)
Scaling, $\sigma_{\gamma^* p} = \sigma_{\gamma^* p} (\eta)$, in $\eta$
yields $\sigma_{\gamma^* p} / \sigma_{\gamma p} \rightarrow 1$ for
$W^2 \rightarrow \infty$ at fixed $Q^2$ and provides evidence for saturation.

\noindent
ii)
Sum rules relate the elastic component in diffractive production to the total
cross section, the terminology GVD/CDP being appropriate for low-x DIS.

\noindent
iii)
The excess of diffractive production over the elastic $(q \bar q)$ component
is presumably due to higher spin components, and accordingly

\noindent
iv)
any theory of diffraction has to discriminate between an inelastic and an elastic
component and must be examined with respect to its compatability with the
total cross section, $\sigma_{\gamma^* p}$.


\end{document}